\documentclass[12pt]{article}
\usepackage[totalheight = 23cm, totalwidth = 17cm]{geometry}

\usepackage{amsmath,amssymb,graphicx}
\usepackage{epsf}
\usepackage{epstopdf}
\usepackage {amssymb}
\usepackage{color}
\newcommand{\nc}{\newcommand}
\nc{\ba}{\begin{eqnarray}}
\nc{\ea}{\end{eqnarray}}
\newcommand\be{\begin{equation}}
\newcommand\sub[1]{_{\rm #1}}
\newcommand{\dchil}{\delta\chi\sub L}
\newcommand{\zetal}{\zeta\sub L}
\newcommand{\zetas}{\zeta\sub S}

\newcommand{\fnl}{f\sub{NL}}
\newcommand\ee{\end{equation}}

\newcommand\bfx{{\mathbf x}}
\newcommand\bfk{{\mathbf k}}

\newcommand\fNL{{f_{\rm NL}}}

\newcommand{\calR}{{\cal{R}}}
\newcommand{\calP}{{\cal{P}}}

\newcommand{\dchis}{\delta\chi\sub S}
\nc{\x}{{\bf{x}}}

\begin{document}

\begin{titlepage}


\begin{center}
\vskip 1.0cm

{\bf CMB Hemispherical Asymmetry:   
Long Mode Modulation and non-Gaussianity}
\vskip 1cm

{ Mohammad Hossein Namjoo $^a$,  Ali Akbar Abolhasani$^b$, Shant Baghram$^a$, Hassan Firouzjahi$^a$  }

\vskip 0.5cm

\small{\it 
$^{a}$ School of Astronomy, Institute for Research in Fundamental Sciences (IPM) \\
P.~O.~Box 19395-5531, Tehran, Iran
\\
$^{b}$ School of Physics, Institute for Research in Fundamental Sciences (IPM) \\
P.~O.~Box 19395-5531, Tehran, Iran
}

\vskip 1.2cm

\end{center}

\begin{abstract}

The  observed hemispherical asymmetry in CMB map can be explained by modulation from a long wavelength super horizon mode which non-linearly couples to the CMB modes. We address the criticism in [1] about the role of non-Gaussianities in squeezed and equilateral configurations in generating hemispherical asymmetry from the long mode modulation. We stress that the modulation is sensitive to the non-Gaussianity in the squeezed limit.   In addition, we demonstrate the validity of our approach in providing a consistency condition relating the amplitude of dipole asymmetry to $f_{NL}$ in 
the squeezed limit.

\end{abstract}

\end{titlepage}

\setcounter{page}{0}
\newpage
\setcounter{page}{1}


\section{Introduction}
There are several large scale Cosmic Microwave Background (CMB) anomalies  persist to survive even after the Planck's accurate observations \cite{Ade:2013nlj,Ade:2013zuv}. In particular, there seems to exist  a hemispherical asymmetry on the CMB power spectrum \cite{Eriksen:2007pc,Hansen:2008ym, Hoftuft:2009rq, Akrami:2014eta} consistent with the following anisotropic temperature fluctuations  \cite{Gordon:2006ag}
\begin{equation} 
\Delta T(\hat {\bf{n}}) = \Delta T_\text{iso} (\hat{\bf{n}} )\left( 1+A \hat{\bf{p}}.\hat{\bf{n}} \right) \, ,
\end{equation}
in which $\hat{\bf{p}}$ is the preferred direction in the sky, $\hat{\bf{n}}$ is the direction of the look 
to CMB and  $A$ is the amplitude of asymmetry with the best fit value $A=0.07$ for large scales, i.e. $\ell \lesssim 64$. This asymmetry, however, seems to vanish at larger scales, i.e. at scales with $\ell \gtrsim 600$,  indicating a non-trivial scale-dependent dipole asymmetry \cite{Hirata:2009ar, Flender:2013jja}.
This type of asymmetry can be modeled by a modulation of the primordial curvature perturbation power spectrum 
\begin{equation}
\label{P-asym}
\calP^{1/2}_\calR(k,{\bf{x}}) = \left[ 1+ A(k) \frac{\hat{\bf{p}}.{\bf{x}}_\text{cmb}}{x_\text{cmb}} \right] \calP^{1/2}_{\calR\, \mathrm{iso}}(k) \, ,
\end{equation}
where ${\bf{x_{cmb}}}$ is the comoving distance to the surface of last scattering and $\calP_{\calR\, \mathrm{iso}}(k) $ is the isotropic power spectrum.

The most promising method to generate such dipole asymmetry  is the modulation of small scale power spectrum by a long wavelength super horizon mode, namely the Grishchuk-Zel'dovich effect \cite{Grishchuk1978}. In this proposal, a mode with wavelength larger than the Hubble horizon modulates the CMB scale modes leaving observable imprints on CMB power spectrum. This kind of modulation can lead to an asymmetry as studied in \cite {aniso-longmode,  Erickcek:2009at, Lyth-dipole, many}. In \cite{Namjoo:2013fka} we have shown that such a modulation is due to a non-linear correlation between the long mode and small CMB modes, i.e. the squeezed limit non-Gaussianity. This logic was extended to incorporate tensor perturbations in \cite{Abolhasani:2013vaa}. There have been works to obtain scale-dependent asymmetry either by using  isocurvature perturbation or non-Bunch-Davies initial state \cite{Erickcek:2009at,Firouzjahi:2014mwa}.

Recently the paper \cite{Lyth:2014mga} appeared in which the author criticized our approach  employed in \cite{Namjoo:2013fka, Abolhasani:2013vaa} to obtain a consistency relation between the amplitude of dipole asymmetry and the squeezed limit non-Gaussianity. In this note  we show that there is no inconsistency in \cite{Namjoo:2013fka, Abolhasani:2013vaa} and the results in \cite{Namjoo:2013fka, Abolhasani:2013vaa} make sense both mathematically and conceptually. In turn, we also comment on the logic of \cite{Lyth:2014mga} in relating the observed dipole asymmetry 
 to equilateral configurations.
 
\section{Hemispherical asymmetry and non-Gaussianity}
\label{sec1}

Let us now review our approach in \cite{Namjoo:2013fka, Abolhasani:2013vaa} in providing the relation between the squeezed limit non-Gaussianity and the amplitude of dipole asymmetry. In doing so we also  answer the criticisms raised in \cite{Lyth:2014mga} and confirm our approach in \cite{Namjoo:2013fka, Abolhasani:2013vaa}.

Before going through the mathematical presentation let us discuss what one may expect from the relation between dipole asymmetry and non-Gaussianity.
 As discussed before, we assume that  a long wavelength mode (super-horizon mode) is responsible for generating the CMB dipole asymmetry. This means that the CMB-scale observer has to see an effect from the long mode. Obviously this can not happen at linear regime in perturbation theory since different scales can not share information at linear order.{\footnote {
Of course, note that the long mode can affect $a_{\ell m's}$ even at linear order especially at low $\ell$ regime and there are several observational constraints that have to be satisfied by this effect. See e.g. \cite {aniso-longmode,  Erickcek:2009at, Lyth-dipole, many} for discussions in this regard.}} Indeed, any correlation between the long mode and the CMB modes is because of the non-linearity, or practically non-Gaussianity. Since the scales are quite different we expect that such a correlation should be sensitive to the non-Gaussianity at the squeezed limit in which one mode has much larger wavelength than the other two  in the three point correlation function. This is actually what we will justify below in contrast to  \cite{Lyth:2014mga} in which it is claimed that the equilateral configuration has to be considered. 

To start, let us first define the bispectrum of curvature perturbation $\calR$ by 
\ba 
\langle \calR_{\bfk_1} \calR_{\bfk_2} \calR_{\bfk_3}  \rangle = (2 \pi)^3 \delta^3( {\bf k_1+k_2 + k_3})
B_\calR({\bf k_1,k_2 , k_3}) \, .
\ea 
The squeezed limit bispectrum, the bispectrum in the configuration in which $k_1  \ll k_2 \sim k_3 =k$, can be parametrized by
\ba 
\label{squeezedfnl}
B_\calR ({\bf k_1,k , k}) \rightarrow \dfrac{12}{5}\fnl(k_1,k,k) P_{k_1} P_{k} \, ,
\ea 
where $P_k$ is the two point correlation of curvature perturbation and $\fnl$ is the amplitude of squeezed limit non-Gaussianity which can be either scale-dependent or free of scale. The so-called local non-Gaussianity is the case in which $\fnl$ is scale-free. However, we  do not need to restrict ourselves to the local shape. 

Now we make an important assumption and focus  on single source models of early universe. By single source we mean that only one scalar field, say $\sigma$, contributes to the curvature perturbations. There may be more than one field in the system but fields other than $\sigma$ do not contribute in curvature perturbations, they contribute only  to the background expansion.  In this case, one can go to the comoving gauge $\delta \sigma=0$ such that the comoving curvature perturbation $\calR = \psi + H \delta \sigma /\dot \sigma   $ is equal to $\psi$, the curvature perturbation in three-dimensional hyper surface. 
 With this definition the standard curvaton scenario in which all curvature perturbations are generated from the curvaton decay is within our category of single source.

Now we are ready to provide our proof for the explicit relation between dipole asymmetry and squeezed non-Gaussianity. We denote the long super-horizon mode by $\bfk_L$ while the smaller CMB-scale modes are denoted by $\bfk_2= \bfk_3 =\bfk$. 
First of all, note that the effect of a long wavelength mode on small scale modes is just to rescale the background. This is because the  small scale observer can not probe the wave-like nature of the large scale fluctuations. Hence, in the squeezed limit one can write 
\ba 
\label{modulation}
\langle \calR_{\bfk_L} \calR_{\bfk_2} \calR_{\bfk_3}  \rangle \simeq \langle \calR_{\bf k_L} \langle \calR_{\bf k_2} \calR_{\bf k_3} \rangle_{\calR_{\bf k_L}}\rangle 
\simeq 
 P_{\bfk_L} \dfrac{d P_{\bfk}}{d\calR_{\bfk_L}} \, ,
\ea 
where the last equality has been obtained by Taylor expanding around the background in the absence of the long mode. The above relation shows that the squeezed limit bispectrum can modify the small scale power spectrum. On the other hand, from Eq. \eqref{P-asym} one obtains
\ba 
\dfrac{\nabla P_k}{P_k} = \dfrac{2 A(k) \hat {\bf p}}{x_{\mathrm{cmb}}}.
\ea 
Combining this with Eqs. \eqref{squeezedfnl} and \eqref{modulation} one readily obtains
\ba 
\label{consistency}
A(k) \simeq \dfrac{6}{5} \fnl(k_L,k,k) \, k_L \,  x_{\mathrm{cmb}}\,  \calP_{\calR_L}^{1/2}  \, ,
\ea 
where $\calP_{\calR_L} = \calP_{\calR_{k_L}}$ and $\calP_{\calR_k} = k^3 P_k/(2\pi^2) $ is the dimensionless power spectrum per logarithmic momentum interval. 

Eq. (\ref{consistency}) is the consistency relation obtained in \cite{Namjoo:2013fka} which clearly shows that the amplitude of dipole asymmetry is controlled by the squeezed limit non-Gaussianity. Note again that $\fnl$ here can be scale-dependent which can be used to explain the scale-dependent dipole asymmetry as required by the lack of any dipole asymmetry
on $\ell > 600$ \cite{Hirata:2009ar, Flender:2013jja}.  We stress that this relation also gives the consistent results for previously studied models. For single field models in which the Maldacena's non-Gaussianity consistency relation \cite{Maldacena:2002vr, Creminelli:2004yq}
is at work we have $A \sim \fnl \sim 1-n_s$. As a result the level of dipole asymmetry in single source model satisfying Maldacena's consistency condition is too small to produce the observed value of $A$. On the other hand,  for the curvaton model with large enough $f_{NL}$ observable dipole asymmetry can be generated. Note that non-Gaussianity in curvaton model is in local shape which actually peaks at squeezed configuration. It should be mentioned that after imposing the quadrupole constraint \cite {aniso-longmode,  Erickcek:2009at} to find upper bound on the combination $k_L \,  x_{\mathrm{cmb}}\,  \calP_{\calR_L}^{1/2}$ one concludes that \cite{Lyth-dipole}  $A \lesssim 0.01 f_{NL}^{1/2}$ .

The key assumptions in obtaining the consistency relation Eq. (\ref{consistency}) are that we consider  single source models and that the effect of the long mode is just to rescale the background for small scale modes. The latter is the basic assumption for the validity of $\delta  N$ formalism  \cite{Sasaki:1995aw, Sasaki:1998ug, Lyth:2004gb}.  So if $\delta N$ formalism works for a single source model then the consistency  relation (\ref{consistency}) also works. However,  the validity of Eqs. (\ref{modulation}) and
(\ref{consistency}) are more general than $\delta N$ formalism. The reason is that in order for $\delta N$ formalism to work one has to assume that all three modes are super-horizon. However, in deriving Eq. (\ref{consistency}), we have used a less restrictive assumption, we only need that the long mode $k_L$ to
be super-horizon while the modes $k_2 \simeq k_3$ can be inside or near horizon crossing. 

The main criticism in \cite{Lyth:2014mga} is that our consistency condition Eq. (\ref{consistency}) does not work for curvaton-type model. It is argued that \cite{Lyth:2014mga} for curvaton model the above approach, starting with Eq. (\ref{modulation}),    would yield $f_{NL} \sim  n_s -1$ as in Maldacena's consistency condition predicting far too small value of $A$ to explain the observed dipole asymmetry. 
In next Section we explicitly calculate $f_{NL}$ for curvaton model using the method employed in Eq. (\ref{modulation}) and obtain the well-known result for $\fnl$ in curvaton model. This confirms the validity of 
Eq. (\ref{modulation}) as long as we work with single source scenarios such as in curvaton model.

Here we also clarify one potential point of confusion.  In the above calculations we did not need to obtain $\fnl$ from Eqs. (\ref{modulation}) or (\ref{consistency}). We assume that there exists a non-linearity at squeezed limit and one calculates $\fnl$ by other methods such as from in-in or $\delta N$ formalisms. Then the consistency relation Eq. (\ref{consistency})  relates this non-linearity to dipole asymmetry on small scale perturbations.  Note 
even if one is interested in generating scale-dependent  asymmetry from 
scale-dependent non-Gaussianity our approach is still valid. We will discuss more about this issue later on.

Before ending this Section it is worth to mention that the above relations clearly show  that one can not relate non-Gaussianity in other configurations, e.g. equilateral configurations considered in \cite{Lyth:2014mga}, to dipole asymmetry. 


\section{Squeezed limit non-Gaussianity and background modulation}

As discussed before it is argued in  \cite{Lyth:2014mga} is that our consistency condition Eq. (\ref{consistency}) does not work for curvaton-type model. It is argued in  \cite{Lyth:2014mga} 
that the starting point Eq. (\ref{modulation})  yields the wrong result  $f_{NL} \sim  n_s -1$ for curvaton model. Here we answer this criticism and explicitly demonstrate that starting from Eq. (\ref{modulation}) indeed we obtain the well-known result for $\fnl$ in curvaton model. 

Combining Eqs. (\ref{modulation}) and  (\ref{squeezedfnl}) we have 
\ba
\label{rescale}
\langle \calR_{k_L} \langle \calR_{k_1} \calR_{k_2} \rangle_{\calR_{k_L}}\rangle =
\dfrac{12}{5} f_{NL} P_{k_L} P_{k_2} .
\ea
We would like to check if Eq. (\ref{rescale}) gives the correct formula for $f_{NL}$ for single source 
models such as the curvaton scenario. Taylor expanding the left hand side  of \eqref{rescale} yields
\ba 
\dfrac{12}{5} f_{NL}  = \dfrac{d\ln \calP_\calR}{d \calR_L} \, .
\ea 
Therefore, in order to calculate $\fnl$ we have to find the imprints of $\calR_L$ in $\calP_\calR$.
Since we work with single source model, as discussed before, one can trade off the comoving curvature perturbation $\calR$ with the three-dimensional curvature perturbation $\psi$ in comoving gauge by
$\calR = \psi$. As a result the effects of  long mode $\calR_L$ in metric can be incorporated as follows
\ba 
ds^2 = -dt^2 + a^2 e^{2\calR_L} d\bfx^2.
\ea 
The key observation is that  $\calR_L$ is not necessarily conserved for single source models on super-horizon scales as in curvaton model. A similar situation also arises in non-attractor models \cite{nonattractor} in which the curvature perturbation is not frozen on super-horizon scales.  Therefore, unlike \cite{Creminelli:2004yq, Maldacena:2002vr},
 we can not remove $\calR_L$ by rescaling the spatial coordinates. Instead, we can absorb it into the scale factor since the $x$-dependence of the super-horizon mode can be neglected. 
 As a result, in the presence of the long mode the scale factor modifies to
\ba 
\tilde{a} = a e^{\calR_L} \, .
\ea 
In the spirit this is very similar to $\delta N$ formalism \cite{Lyth:2004gb}. 

Changing the variable from $\calR_L$ to $\tilde{a}$ and replacing $\tilde{a}$ with 
$a$ for simplicity one has 
\ba 
\label{fNL}
\dfrac{12}{5} f_{NL}  = \dfrac{d\ln \calP_\calR}{d \ln a}=\dfrac{d\ln \calP_\calR}{d N}.
\ea 
This is an interesting result. Firstly, note that it gives the Maldacena's consistency relation for single field models of inflation in which $\calR$ is frozen on super-horizon scales. 
In this case, we note that the power spectrum on super-horizon scale is given by 
\ba 
\calP_\calR ={ \calP_{\calR0}} \left(\dfrac{k}{a H} \right)^{n_s-1}
\ea  
where ${ \calP_{\calR0}}$ is the power spectrum at the time of horizon crossing. Noting that the Hubble parameter $H$ is nearly constant during inflation one concludes $d\calP_\calR/d\ln a = -(n_s-1)$ which yields the expected result
\ba
\dfrac{12}{5} f_{NL}= 1-n_s.
\ea
We comment that  the relation \eqref{fNL} is more general than the $\delta N$ formalism, since the latter 
has order slow-roll errors in predicting the value of $f_{NL}$ which comes from the near horizon crossing effects. 

Now consider a general single source model in which the $\delta N$ formalism is at work. In this case we have 
\ba 
\calP_\calR= N_\sigma^2 \calP_{\delta\sigma}
\ea 
in which $N_\sigma$ indicates the derivative of number of e-folds with respect to the field $\sigma$ which
generates the curvature perturbation.  If $\delta \sigma$ is massless during inflation its 
power spectrum  $\calP_{\delta\sigma}$ is constant in time. 

Note that for single source models we have $N=N(\sigma)$. As a result, assuming  $\calP_{\delta\sigma}$ is nearly constant, from Eq. (\ref{fNL}) we obtain
\ba 
\label{fNL2}
\dfrac{6}{5} f_{NL} = \dfrac{N_{\sigma\sigma}}{N_{\sigma}^2}  \, .
\ea 
Interestingly, this is in exact agreement  with the results obtained from non-linear $\delta  N$ formalism. Therefore, it is expected that our results  Eqs. (\ref{fNL}) and (\ref{fNL2}) should yield the correct result for
the curvaton model. Below we demonstrate  this explicitly.

At linear order for curvaton model we have 
\ba 
\calR=\dfrac{2r}{3\sigma} \delta \sigma
\ea 
in which $\sigma$ is the value of the background curvaton field at initial time of oscillation and the parameter $r$ is related to the curvaton and radiation energy densities, $\rho_\sigma$ and $\rho_\gamma$,  at the time of decay via
\ba
\label{r-def}
r= \frac{3 \rho_\sigma}{ 3 \rho_\sigma + 4 \rho_\gamma} \bigg|_{\mathrm{decay}} \, .
\ea
This yields
\ba
  \calP_\calR = \dfrac{4 r^2}{2 \sigma^2}  \calP_{\delta\sigma} \, .
\ea
In calculating ${d\ln \calP_\calR}/{d N}$  extra care has to be taken. The reason is that the above power spectrum has been calculated at the hyper-surface of curvaton decay  where the total energy is a constant. Hence when we change the number of e-folds we need to change $\sigma$ field at the same time to compensate the change in the energy density by changing the number of e-folds. In other words we have 
\ba 
\calP_\calR = \calP_\calR(N,\sigma(N)) \, .
\ea 
One can quantify the above discussion by writing the total energy density as 
\ba 
\rho_{\mathrm{tot}}  = \dfrac{1}{2}m_\sigma^2 \sigma^2 e^{-3N} + \rho_{0\gamma} e^{-4N}
\ea 
in which $\rho_{0\gamma}$ represents the initial value of radiation energy density.  
Noting that  $\rho_{\mathrm{tot}}$ is constant on the curvaton decay hyper-surface and neglecting the perturbations from inflaton field (as we consider the single source model) 
we obtain the following relation
\ba 
\dfrac{d\sigma}{dN} = \dfrac{3\sigma}{2r}.
\ea 
We then have 
\ba 
\dfrac{12}{5} f_{NL} = \dfrac{\partial \ln \calP_\calR}{\partial N} + \dfrac{\partial \ln \calP_\calR}{\partial \sigma} \dfrac{d\sigma}{dN} \, .
\ea 
Now using 
\ba 
\dfrac{\partial \ln \calP_\calR}{\partial N}=2(1-r)
\\
 \dfrac{\partial \ln \calP_\calR}{\partial \sigma} \dfrac{d\sigma}{dN} = \dfrac{3}{r} - 6 \, ,
\ea 
we obtain 
\ba 
f_{NL} = \dfrac{5}{4r} -\dfrac{5}{3} -\dfrac{5r}{6} \, ,
\ea 
which is in exact agreement with the known results for $\fnl$ in curvaton model \cite{Lyth:2005fi, Sasaki:2006kq}. 

This provides a non-trivial support for the validity of our approach in \cite{Namjoo:2013fka} in treating $k_L$
as a modification of the background for small scales $k_1$ and $k_2$ and in using 
Eq. (\ref{modulation}),  or equivalently Eq. (\ref{rescale}),  as a starting point to obtain the consistency relation (\ref{consistency}). Having this said, we stress again that our aim in \cite{Namjoo:2013fka}  was not to use Eq. (\ref{rescale}) to calculate $f_{NL}$. Instead we have used
Eq. (\ref{rescale}) to obtain Eq. (\ref{consistency}) as a relation between the amplitude of dipole asymmetry and the squeezed non-Gaussianity. It is assumed that one uses an independent method, such as the in-in or $\delta N$ formalisms, to actually calculate the value of $\fnl$.

\section{Comments on  \cite{Lyth:2014mga}}

In \cite{Lyth:2014mga}  Lyth has studied the possible link between a scale-dependent asymmetry and
the question of shortage of power on low $\ell$. This is an  interesting idea to see
whether these two anomalies in CMB map are linked together. Having this said, we comment on \cite{Lyth:2014mga} about
the role of squeezed and equilateral non-Gaussianities in generating scale-dependent asymmetry and explaining the low-$\ell$ power shortage. Specifically, as we discussed consistently above, it is  the squeezed limit non-Gaussianity which is relevant for dipole asymmetry and not other configurations.

Let us now have a look on the key relation in \cite{Lyth:2014mga}
\ba
\zeta(\bfx) &=&  N(\chi(\bfx)) - N(\chi_0) \\
&=& N'(\chi_0)) \left( \dchis(\bfx) + \dchil(\bfx) \right)
+ \frac12 N''(\chi_0)) \left( \dchis(\bfx) + \dchil(\bfx) \right)^2
+\cdots \\
\label{deltaN}
&\equiv& \left( \zetas(\bfx) + \zetal(\bfx) \right)
 + \frac35\fnl(k)
\left( \zetas(\bfx) + \zetal(\bfx) \right)^2 + \cdots
\label{fnlk}
 \\
&=& \zetas(\bfx) + \frac35\fnl(k)  \zetas^2(\bfx)
 + \frac65\fnl(k) \zetal(\bfx) \zetas(\bfx) \nonumber \\
&& +\zetal(\bfx) + \frac35 \fnl(k)  \zetal^2(\bfx) +\cdots
, \label{final} 
\ea
in which, following the notation in  \cite{Lyth:2014mga}, $\chi$ is the curvaton field, $\zeta$ is the curvature perturbation on constant energy surface, $\delta \chi_S$ is the small scale fluctuation while $\delta \chi_L$
is the long mode fluctuation. 

We note that $\fnl$ here is the quasi local non-Gaussianity calculated at {\it equilateral configuration} \cite{Byrnes:2009pe,Byrnes:2010ft} . The quasi local non-Gaussianity can be obtained if the local non-Gaussianity depends on scale. However the above formalism is valid for such non-Gaussianity only if the running of $\fnl$ as a function of scale is weak enough which might not be the case for the aim of scale-dependent asymmetry. If the scale-dependence is large it is not straightforward to obtain an expression similar to  $\delta N$ expansion in real space.  Indeed, in the above formalism, the scale-dependence of $\fnl$ in Fourier transformation has not been considered  and $\fnl$ only implicitly depends on scale when it has been computed at the time of horizon crossing. 
Roughly speaking, we require the spatial variation of $\fNL$ to be sufficiently small within the Hubble patch, since the minimum required volume to perform Fourier transformation is comparable to the Hubble scale. If $\fnl$ has a non-trivial scale-dependence, as it is suggested from the scale-dependence of asymmetry, then the above formalism in real space has to be revisited. 
Instead, one may work with non-local extension of the above local form of non-linearity. A more appropriate formalism for this aim is to use the kernels defined in \cite{Scoccimarro:2011pz} or to work directly in Fourier space, as we did in our approach.

As we mentioned before $\fnl$ in the above relation has been computed at equilateral configuration. This, however, is in tension  with \eqref{fnlk} in which the author decomposed the curvature perturbation into small and long wavelength modes. This decomposition actually shows  that there is a correlation between the long and short modes resulting in an asymmetry which is in tension with the assumption $k_1=k_2=k_3=k$.   

Perhaps the reason that has led  to the above  treatment is the idea in \cite{Lyth:2014mga} to obtain scale-dependent asymmetry and power-deficit by considering a scale-dependent $\fnl$. 
This goal can be consistently embedded in our approach. Eq. \eqref{consistency} shows that if the squeezed limit non-Gaussianity depends on scale then the asymmetry will also be scale-dependent. In order for asymmetry to vanish at scales $\ell \gtrsim 64$ one requires 
\ba 
\fnl (k_L,k,k) \to 0 \quad \, \mathrm {for } \, \, k^{-1}\lesssim  \frac{x_{\mathrm{cmb}} }{64}
\ea   
for fixed value of $k_L$. Note that for a fixed $k_L$ we still can change the other two momenta in the triangle.  The reason that we fix $k_L$ is that we consider one long mode which correlates with all CMB modes. 
 
 In \cite{Lyth:2014mga} the author employed the curvaton model as a candidate for obtaining both 
 scale-dependent asymmetry and power deficit. However, it is generally difficult to obtain such 
 scale-dependence in curvaton model \cite{Erickcek:2009at}. Models based on non-attractor models \cite{nonattractor} or models with non-Bunch-Davies 
 initial conditions as studied in \cite{Firouzjahi:2014mwa} may have better chances to fulfill this job. 
 In non-attractor models the curvature perturbation evolves for few e-folds on super-horizon scales and
 large $\fnl$ can be obtained. Once the system reaches the attractor regime $\calR$ freezes and 
 one obtains the usual formula $\fnl \sim n_s -1$. Therefore, for scales which are outside the horizon during
 the non-attractor phase one can obtain large dipole asymmetry. One can arrange that the sub-CMB modes
 leaves the horizon during the attractor phase with negligible $\fnl$ so a scale-dependent $\fnl$ and 
 $A$ is automatically generated in non-attractor models. As for models with non-Bunch-Davies  initial condition,  it has been shown in \cite{Firouzjahi:2014mwa} that one can obtain scale-dependent asymmetry from non-Bunch-Davies initial state. In addition, the model has enough free parameters which may also address the shortage of power on low $\ell$.  \\
 
 In summary, in this note we have emphasized the role of squeezed limit non-Gaussianity in generating hemispherical asymmetry. In order for the long super-horizon mode to affect the small CMB-scale modes, the non-linearity should be in the form of squeezed limit and not other configurations. In response to the criticism raised in \cite{Lyth:2014mga} we have explicitly demonstrated that our approach in treating the long mode as a modification of background for the other two
 small modes correctly reproduces the known result for $\fnl$ in curvaton scenario. 
 
 The possible link between dipole asymmetry and the shortage of power on low $\ell$ is an
 interesting idea as studied in \cite{Lyth:2014mga}. It is not easy to generate scale-dependent power
 asymmetry in curvaton \cite{Erickcek:2009at}. Therefore, it is an interesting open question  to relate dipole asymmetry and the shortage of power in low $\ell$ by allowing the squeezed limit $\fnl$ to have non-trivial 
 scale-dependence. \\

{}

\end{document}